# Influence of non-Newtonian rheology on magma degassing

Thibaut Divoux,[1] Valérie Vidal,[1] Maurizio Ripepe,[2]

and Jean-Christophe Géminard[1]

[1]Laboratoire de Physique - CNRS, Ecole Normale Supérieure de Lyon, 46 Allée d'Italie, 69364 Lyon cedex 07, France.

[2]Dipartimento di Scienze della Terra, Università degli Studi di Firenze, via La Pira, 4 - 50121 Firenze, Italy.

**Abstract.** Many volcanoes exhibit temporal changes in their degassing process, from rapid gas puffing to lava fountaining and long-lasting quiescent passive degassing periods. This range of behaviors has been explained in terms of changes in gas flux and/or magma input rate. We report here a simple laboratory experiment which shows that the non-Newtonian rheology of magma can be responsible, alone, for such intriguing behavior, even in a stationary gas flux regime. We inject a constant gas flow-rate $Q$ at the bottom of a non-Newtonian fluid column, and demonstrate the existence of a critical flow rate $Q^*$ above which the system spontaneously alternates between a bubbling and a channeling regime, where a gas channel crosses the entire fluid column. The threshold $Q^*$ depends on the fluid rheological properties which are controlled, in particular, by the gas volume fraction (or void fraction) $\phi$. When $\phi$ increases, $Q^*$ decreases and the degassing regime changes. Non-Newtonian properties of magma might therefore play a crucial role in volcanic eruption dynamics.

## 1. Introduction

Among the different factors controlling the eruptive dynamics of volcanoes, degassing processes are thought to directly control the intensity and style of the explosive activity [*Ripepe*, 1996; *Parfitt*, 2004; *Houghton and Gonnermann*, 2008]. Due to decompression during magma ascent, bubbles nucleate, grow and coalesce. This two-phase flow exhibits different regimes: bubbly flow, where the liquid is filled with a dispersion of small bubbles; slug flow, where larger bubbles (*slugs*), with a diameter of the order of the conduit size, rise quasi-periodically; churn flow, where slugs break down, leading to an oscillatory regime; and annular flow, where the gas flows continuously at the conduit center and the liquid is confined on the walls [*Wallis*, 1969; *Taitel et al.*, 1980]. Although these regimes are well characterized in the engineering literature, their application to volcanoes, where the degassing activity can change in time [*Gonnermann and Manga*, 2007] or several mechanisms of degassing can co-exist at once [*Houghton and Gonnermann*, 2008], still raises many questions.

Temporal changes in the degassing process are observed through the alternation between these different regimes. In particular, Strombolian activity has been mathematically described as an intermittent phenomenon [*Bottiglieri et al.*, 2008], where intermittency here refers to a system (the volcano) which can switch back and forth between two qualitatively different behaviors, while all the control parameters remain constant. Non-linearity is essential to describe and reproduce spontaneous changes of activity observed in volcanic eruptions [*Woods and Koyaguchi*, 1994; *Melnik and Sparks*, 1999; *Bottiglieri et al.*, 2008], but its origin still remains under debate. Most of the previous studies invoke changes in gas flux and/or magma flow rate at depth [*Ripepe et al.*, 2002, *Woods and Koyaguchi*, 1994].

Bubbles rising in magma can also lead, by deformation under shear and coalescence, to the formation of channel-like bubble networks, which makes it possible for the gas to percolate through the system [*Okumura et al.*, 2006; *Burton et al.*, 2007; *Okumura et al.*, 2008]. If these structures provide a possible explanation for quiescent degassing, they do not explain, however, the intermittence observed in the degassing regimes.

## 2. Magma Rheology

At low strain rate and small volume fraction of crystals, the magma can be considered as Newtonian. Departure from this behavior may be observed on the one hand, if the magma has a crystal fraction $\phi_c > 0.1$ [*Fernandez et al.*, 1997] as, for example, Strombolian magma [*Francalanci et al.*, 1989]. On the other hand, for strain rates higher than $1/\tau$, where $\tau$ is the relaxation timescale, magma exhibits brittle behavior [e.g. *Dingwell*, 1996]; for fully molten silicate systems, for instance, the transition to non-Newtonian behavior seems to occur at strain rates 2-3 orders of magnitude lower than $1/\tau$ [*Webb and Dingwell*, 1990]. One characteristic of the non-Newtonien behavior is that the magma viscosity decreases with increasing strain rate [*Gonnermann and Manga*, 2007, *Webb and Dingwell*, 1990, *Lavallée et al.*, 2007, *Caricchi et al.*, 2007]. Moreover, for higher crystal content ($\phi_c > 0.3$) magma seems characterized by the presence of a yield strength [*Fernandez et al.*, 1997].

Contrary to viscous Newtonian fluids, fluids exhibiting a yield strength are able to trap bubbles permanently. Indeed, in Newtonian fluids, small bubbles may rise very slowly but at long time, they are always able to reach the free surface. In fluids characterized by the existence of a yield strength $\sigma_c$, bubbles with a radius $r<3\sigma_c/(4\Delta\rho g)$, where $\Delta\rho$ the difference between the gas and fluid density and *g* the gravitational acceleration, cannot rise and remain trapped in the fluid. The presence of small bubbles can strongly modify the rheological properties of magma [*Manga et al.*, 1998, *Manga and Loewenberg*, 2001]. Moreover, the rheological properties of magma can evolve through time with temperature, volume fraction of bubbles, crystal content, etc. This point led several authors to propose that the variations of magma rheology could be responsible for the spontaneous changes in eruptive styles observed at different volcanoes [*Jaupart and Vergniolle,* 1988; *Dingwell*, 1996, *Caricchi et al., 2007, Gonnermann and Manga*, 2007].

We propose, based on laboratory experiments, that shear-thinning properties and the existence of a yield strength in magmas are key ingredients which may control the eruptive dynamics. Moreover, the evolution of shear-thinning and yield strength with increasing bubble content is an important factor governing the changes in degassing regimes.

## 3. Experimental setup

The experimental setup consists of a vertical plexiglas tube (inner diameter 74 mm, height 270 mm) filled with a non-Newtonian fluid. The fluid is a diluted solution (15% in mass of distilled water, unless specified) of a commercial hair-dressing gel (*Gel coiffant, fixation extra forte,* Auchan). This fluid can be easily supplied in large quantities, the mixture is reproducible and stable in time if well preserved from drying. The fluid is strongly shear-thinning for shear rates higher than $10^{-2}$ s$^{-1}$, and exhibits a yield strength $\sigma_c$ ~ 40 Pa (see Supplementary Material). A mass-flow controller (Bronkhorst, Mass-Stream series D-5111) injects air inside a tank at a constant flow-rate $Q$ (from $Q_{min}$ = 0.17 cm$^3$/s to $Q_{max} = 1.74$ cm$^3$/s). The tank is partially filled with water and, above the fluid column, a small container also filled with water maintains a saturated humidity level at the gel free surface, preventing any significant drying of the sample during the experiment. Air is injected at the bottom of the fluid column via rigid tubes (typical diameter 8.0 mm) and a nozzle with a diameter of 2.0 mm. We measure the variation $\delta P$ of the overpressure inside the tank with a low-pressure differential-sensor (Honeywell S&C, 176PC28HD2) connected to a multimeter (Keithley, 196). Pressure is recorded over a few days at 5 Hz, with an accuracy of 0.5 Pa. Visual observation is inhibited as the fluid progressively fills up with small bubbles. Monitoring the pressure variations in the tank makes it possible to study the degassing activity. The experiment reproducibility is ensured by the following procedure: the gel solution, initially free of bubbles, is poured inside the tube (initial height $h_0$); The flow-rate is set to the smallest accessible value, $Q_{min}$, and is increased by steps of amplitude $\delta Q$ (~$Q_{min}$), up to the maximum value $Q_{max}$. For each flow-rate $Q$, $\delta P$ is recorded for at least 3 days. Then, we reverse the process by decreasing the flow-rate $Q$ down to $Q_{min}$, still with $\delta Q$ steps. The whole process is here referred to as a *flow-rate cycle*.

At the beginning of the experiment, before the first flow-rate cycle is applied, the fluid is free of bubbles and the void fraction $\phi=0$ (*pure gel*). During the first flow-rate cycle, the successive bubbles bursting at the surface generate small satellite bubbles, which remain trapped in the bulk due to the fluid yield strength. These bubbles progressively accumulate below the free surface [Divoux et al., 2009; Vidal et al., 2009]. In stationary regime, a vertical gradient of small, trapped bubbles is thus observed along the fluid column.

## 4. Variable degassing process at constant flow-rate

First, we focus on the degassing dynamics when injecting a constant flow-rate. In stationary regime, under a critical flow-rate ($Q<Q^*$), cusped bubbles form, rise and burst quasi-periodically at the free surface of the non-Newtonian fluid (*bubbling regime*). The overpressure $\delta P$ exhibits successive rises and drops, each sequence associated with a single bubble formation at the bottom of the fluid column (Fig. 1, top). If the imposed flow-rate is above $Q^*$, we observe that the system spontaneously alternates between two states: the *bubbling* and the *channeling* regimes (Fig. 1, center). We call *channeling* when an open channel connects the nozzle to the fluid free surface. In this regime, the overpressure inside the chamber only differs from the atmospheric pressure by the pressure drop in the channel. Nonetheless, $\delta P$ is characterized by a low frequency evolution due to small changes in the mean channel diameter and still exhibits small amplitude oscillations associated with the upwards advection of the irregular channel shape (Fig. 1, bottom) [*Kliakhandler*, 2002]. .

For each applied flow-rate $Q$, we run the experiment for a long time (typically, a few days) and measure, for each value of $Q$, the percentage of time spent in the bubbling regime. In Figure 2, we report the probability for the system to be in the bubbling regime. The transition between the *bubbling* regime, in which the degassing process occurs through the successive emission of bubbles, and the *intermittent* regime, in which the system alternates between bubbling and channeling, is sharp. It is therefore possible to define clearly a critical flow rate $Q^*$ (Fig. 2). Note that the system dynamics, below and above $Q^*$, is not symmetric. For $Q < Q^*$, the degassing process always corresponds to the bubbling regime, and the formation of the channel is never observed. For $Q > Q^*$, however, the channel always has a finite lifespan and collapses back to the bubbling regime, even for large flow-rate; in this latter case, its collapse is followed almost immediately by the opening of a new channel. We find that the critical flow-rate $Q^*$ does not depend significantly on the initial height $h_0$ of the fluid column and of the volume of the air tank, and only depends on the rheological properties of the fluid [*Divoux et al.*, 2009].

## 5. Gas volume fraction and magma rheology

Here, we focus on the transient regime during which an initially pure gel ($\phi=0$) progressively fills up with small bubbles trapped in the column. We thus consider the increase of the gas volume fraction $\phi$ through time. The satellite bubbles are generated both by the bubbles bursting at the surface, or by the channel pinch-off during the transition between the channeling and bubbling regime. They preferentially accumulate at the center and top of the fluid column, leading to a roughly linear variation of the void fraction as a function of depth inside the fluid column (vertical bubble gradient). It is important to note that contrary to previous works which reported the role of porosity on bubble coalescence, in our system the channel is not formed by bubble coalescence in the bulk. It is rather a Saffman-Taylor finger-like structure [*Saffman and Taylor*, 1958], growing in the fluid from the bottom air injection nozzle up to the fluid free surface. The presence of small bubbles, however, changes the rheological properties of the fluid (see *Divoux et al.* [2009] and Supplementary Material): both the viscosity and the yield strength decrease when the bubble content increases. Investigating a system where the bubble fraction evolves in time is, in this experiment, equivalent to consider a change in rheology through time, the latter being coupled to the system dynamics.

We define $Q^*_1$ as the threshold flux measured during the first flow-rate cycle, starting with a gel initially free of bubbles. During the second cycle, the column has been filled with small bubbles and we get $Q^*_2 < Q^*_1$ (Fig. 2). From the third flow-rate cycle, the small-bubble gradient does not evolve anymore and $Q^*$ remains equal to $Q^*_2$. The critical flow-rate $Q^*$ is therefore linked to the rheological properties of the fluid, which depend here on the gas volume fraction $\phi$. Note that the closer the imposed flow-rate is to $Q^*$, the faster the column fills up with bubbles. Indeed, for $Q \ll Q^*$ or $Q \gg Q^*$, the vertical bubble gradient is not fully developped after one week, while for $Q \sim Q^*$, the bubble gradient is established within a few hours. This observation points out the importance of the channel pinch-off in the satellite bubbles production.

We measure the average gas volume fraction in the system as $\bar{\phi} = \dfrac{h - h_0}{h_0}$, where $h$ is the total column height and $h_0$ its initial height, as a function of the imposed constant flow-rate $Q$ (Fig. 3). During the first cycle, $\bar{\phi}$ exhibits a maximum around $Q^*_1$. Indeed, the closer $Q$ is to $Q^*$, the faster the bubble gradient develops, leading to the column expansion. Note that this maximum exists only for the first cycle ($Q^*_1$). During the following cycles, the bubble gradient is already established and there is no further evolution of the gas volume fraction.

When increasing $Q$, the formation of a dome on the top of the gel column is observed (Fig. 1). We measure the dome height, $\Delta h$, during the first cycle when increasing (Fig. 4, top) and then decreasing (Fig. 4, bottom) $Q$. As the flow-rate reaches $Q^*_1$, at the transition from the bubbling to the channeling regime, the dome collapses. When reducing the flow-rate, we still observe a maximum in the dome height for $Q \sim Q^*_1$.

## 6. Implication for Magma Degassing

The two degassing regimes in this experiment might be interpreted in terms of volcano dynamics. In order to attempt an analogy, however, it is necessary to consider a frame where only degassing occurs – no net mass flux of fluid. In the stationary regime (constant gas flux and fixed magma rheology, i.e. fixed gas volume fraction), a volcano will exhibit, in this frame, two different degassing dynamics: for $Q < Q^*$, it will be in the bubbling regime, exhibiting quasi-periodic bubble bursting, whereas for $Q > Q^*$ it will experience an intermittent behavior, spontaneously alternating between the bubbling and channeling regimes (Fig. 2). If the gas volume fraction evolves through time (transient regime), it will modify the magma rheological properties, and then the critical flow rate $Q^*$, changing the volcano degassing dynamics. For instance, a volcano with an initial small void fraction and constant gas flux $Q$ just below the critical $Q^*$ will be in a persistent quasi-periodic bubbling regime. Due to the increase of the gas volume fraction $\phi$, $Q^*$ will decrease, leading to $Q > Q^*$ and driving the volcano to an unstable intermittent dynamics.

While the bubbling regime has an equivalent in the bubbly flow of volcanoes where bubbles rise and burst at the magma surface, generating quasi-periodic explosions, the channeling regime is more difficult to interpret in terms of magma dynamics. On the one hand, it could represent the annular flow, where the gas continuously flows through the conduit, generating sustained lava fountains. On the other hand, the continuous, quiet emission of gas in this regime recalls more quiescent magma degassing.

## 7. Conclusion

Laboratory experiments of degassing in a non-Newtonian fluid column demonstrate that the rheological properties of the fluid, alone, can be responsible for the observed alternation between different degassing regimes. This result can be interpreted in terms of variability of the explosive activity on volcanoes. This mechanism does not dismiss previous ideas [e.g. *Jaupart and Vergniolle*, 1988; *Melnik and Sparks*, 1999] but rather points out that the non-Newtonian nature of magmas could be responsible for the intermittent degassing observed on active volcanoes. Our experiments indicate that an increase or decrease of explosivity of open-conduit volcanoes could be simply explained as due to the non-Newtonian properties of magma, without any variations of the gas or magma flow rate.

This conclusion is consistent with other theoretical and experimental studies which point out magma rheology as the factor controlling the eruptive dynamics of volcanoes [*Caricchi et al.*, 2007; *Gonnermann and Manga*, 2007, *Lavallée et al.*, 2008]. Besides, we provide an experimental evidence of the formation of a gas-channel different than the one proposed in previous study. In our experiments, the gas channel is not formed due to porosity and bubble coalescence in the bulk, but is rather a finger-like structure, growing from the fluid bottom up to the free surface. We show that, in a shear-thinning non-Newtonian fluid, a gas-channel is likely to form even at constant flow-rate. The formation of these channels favours continuous magma degassing and tends to decrease the volcano explosivity [*Burton et al.*, 2007; *Houghton and Gonnermann*, 2008].

**Acknowledgments**

The authors acknowledge S. Ciliberto and E. Bertin for helpful discussions. T.D. and V.V. are indebted to S. Vergniolle for numerous enlightening discusions, and greatly benefited from the 2008 IAVCEI General Assembly (Reykjavik, Iceland).

**FIGURE CAPTIONS**

**Figure 1.** Pressure variations $\delta P$ vs time $t$ observed at the bottom of the fluid column (inner diameter 74 mm, total height of the cell 270 mm), when injecting air at constant flow-rate $Q$ (see Supplementary Material for the experimental setup). The system exhibits a spontaneous alternation between periodic (bubbles) and continuous (open channel) degassing. Rapid drops of the overpressure mark the emission of successive bubbles at the bottom of the column (top), whereas a slowly oscillating overpressure corresponds to an open channel connecting the injection nozzle to the fluid surface (bottom) [$Q = 0.69$ cm$^3$/s, $h_0 = 7$ cm, $V = 530$ mL].

**Figure 2.** Probability $P_b$ for the system to be in the bubbling regime, vs. flow-rate $Q$. The graph exhibits a sharp transition at a critical flow-rate $Q^*$, which depends on the number of applied flow-rate cycles. We find $Q^*_1 = 0.90$ cm$^3$/s (1$^{st}$ cycle) and $Q^*_2 = 0.63$ cm$^3$/s (2$^{nd}$ cycle). From the third flow-rate cycle, $Q^*$ remains constant ($Q^* = Q^*_2$) [$h_0 = 7$ cm, $V = 530$ mL]. Full (resp. open) symbols correspond to data obtained for increasing (resp. decreasing) flow-rates. Fitting curves (sigmoids) are given as eyeleads.

**Figure 3.** Evolution of the average gas volume fraction $\bar{\phi}$ in the column as a function of the flow-rate *Q*, for the first (top) and second (bottom) cycles. We only represent here the increasing flow rate, and report for both cycles the threshold value $Q^*$ defined in Figure 2 [$h_0$ = 7 cm, *V* = 530 mL].

**Figure 4.** Variation of the dome height *Δh* evolution as a function of the flow-rate *Q* for the first cycle. *Top*: increasing *Q*; *Bottom*: decreasing *Q* [$h_0$ = 7 cm, *V* = 530 mL].

**FIGURE 1**

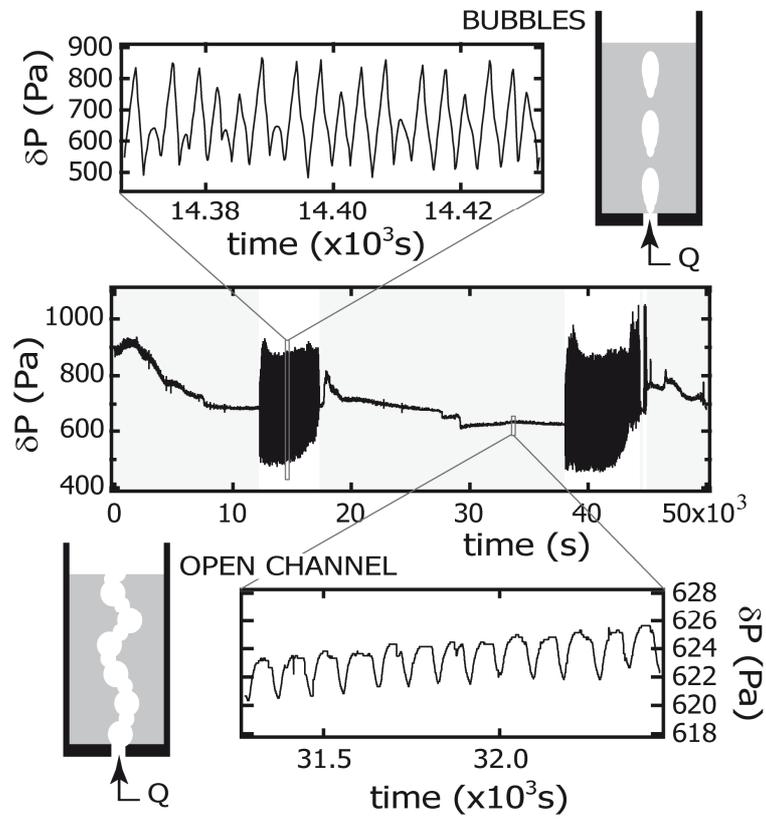

**FIGURE 2**

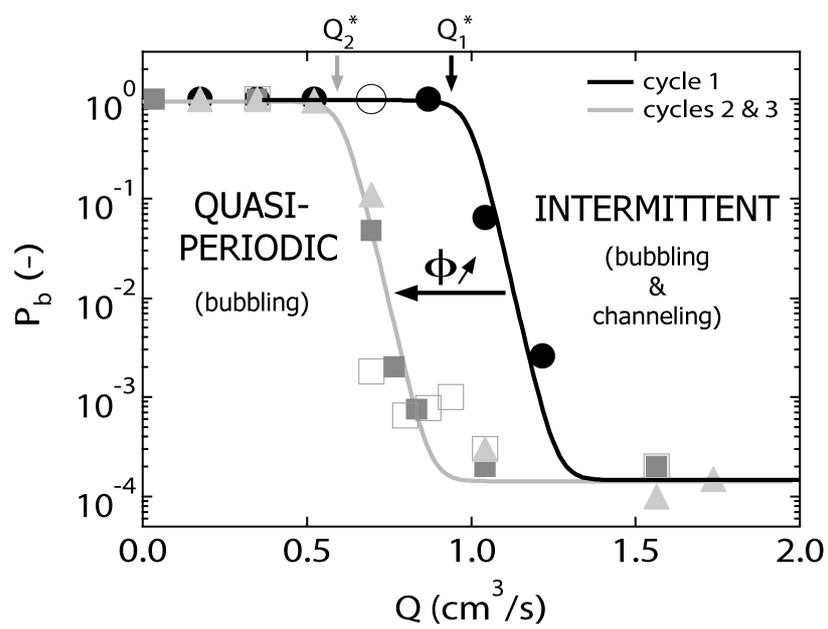

**FIGURE 3**

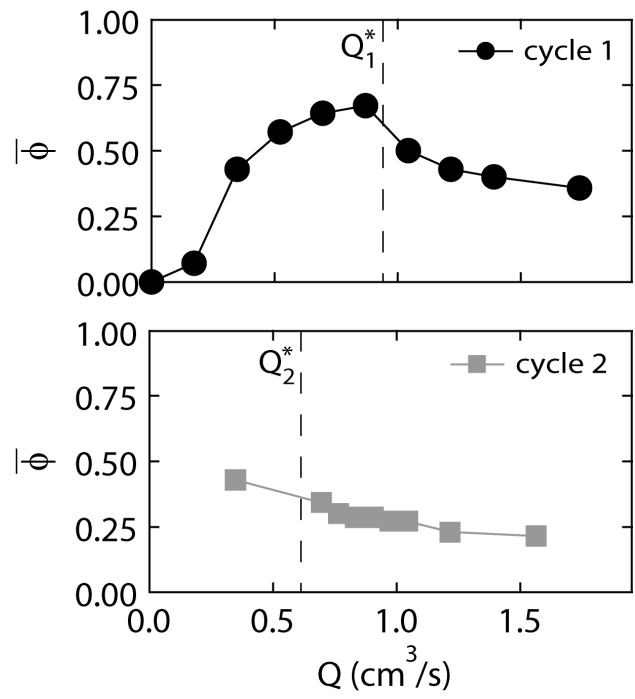

**FIGURE 4**

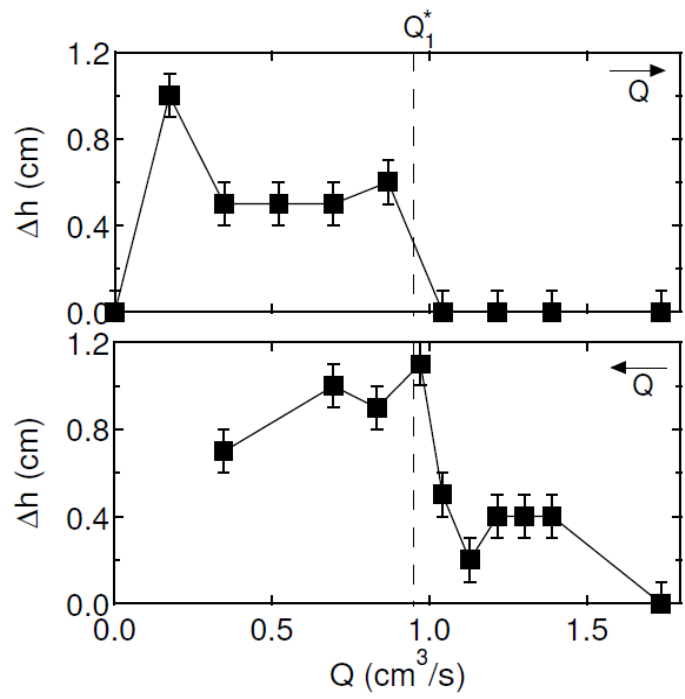

# SUPPLEMENTARY ONLINE MATERIAL

# Influence of non-Newtonian rheology on magma degassing

T. Divoux, V. Vidal, M. Ripepe, and J.-C. Géminard

## 1. Experimental setup
-------------------------------------------------------

The experimental setup is the same than the one described in Divoux et al. [2009]. For sake of clarity, we report it below:

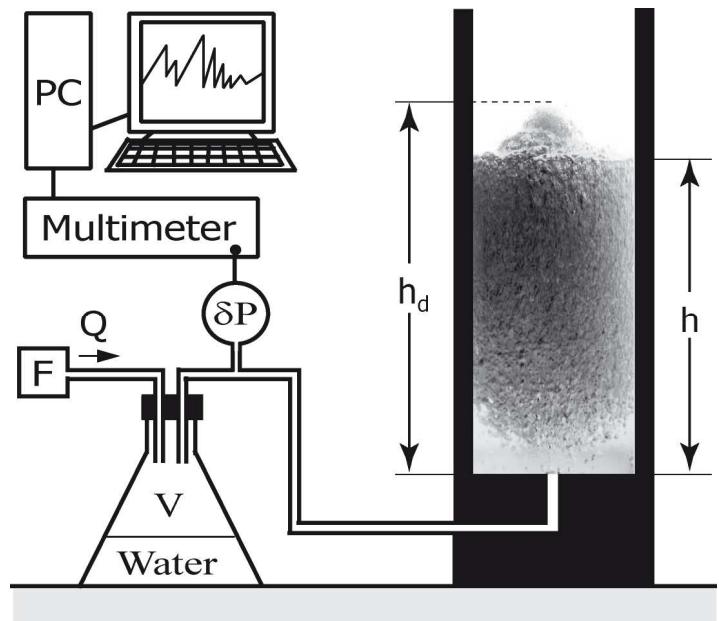

**Figure 1: Sketch of the experimental setup. The fluid column is a real picture from experiments. Note the presence of a vertical gradient of small bubbles trapped in the fluid, as well as a dome at the fluid surface.**

The experimental cell has an inner diameter of 74 mm, and a total height of 270 mm. Air is injected at constant flow-rate $Q$ at the bottom of a non-Newtonian fluid column, via a chamber of volume $V$, through a nozzle of 2 mm diameter. We measure the overpressure variations $\delta P$ in the chamber through time.

We note $h_0$ the initial fluid height, at the beginning of the experiment. After some time, the bubble column fills up with small bubbles trapped due to the fluid yield stress. This vertical bubble gradient is well visible on the picture above. As the void fraction increases, the fluid column height $h$ increases. For a certain range of air flow rate $Q$, we observe the formation of a dome of height $\Delta h = h_d - h$.

## 2. Fluid rheology
-----------------------------------------------------

In the experiment presented in this article, we used, as a non-Newtonian fluid, a diluted solution (15% in mass of distilled water) of a commercial hair-dressing gel (*Gel coiffant, fixation extra forte* Auchan). If preserved from drying, this gel has stable characteristics through time.

The rheological measurements presented here are performed with a rheometer Bohlin Instruments, C-VOR 150, equipped with parallel plates (PP-60, gap 1000 μm). In order to prevent any sliding of the fluid at the walls, sand paper was glued to the plates. All the measurements are performed at constant temperature $T = 25°C$.

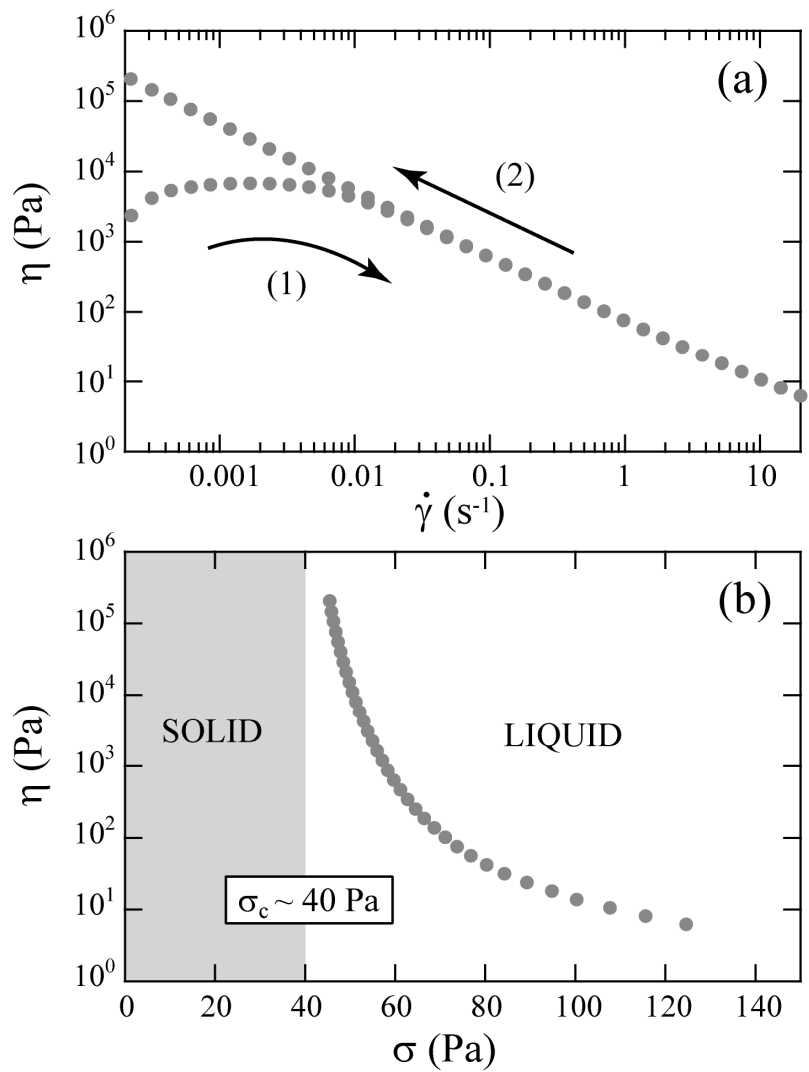

**Figure 2: (a) Viscosity as a function of the applied shear rate. (1) and (2) indicate the measurement cycle (first increasing, then decreasing the shear rate). (b) Viscosity as a function of shear stress for the gel 15%. The gray region indicates the shear stress for which the gel behaves as an elastic solid (for $\sigma<\sigma_c$, where $\sigma_c$ is the yield stress, it does not flow). Only the downward series is represented.**

In Figure 2 are displayed the rheology measurements for the fluid used in this work. The fluid is strongly shear-thinning for shear rates higher than $10^{-2}$ s$^{-1}$ (Fig.2a) and presents a yield stress $\sigma_c \sim 40$ Pa (Fig.2b). In a previous work [Divoux et al., 2009], we mainly used a 10% gel mixture. The graph below presents the change in rheology when varying the gel concentration.

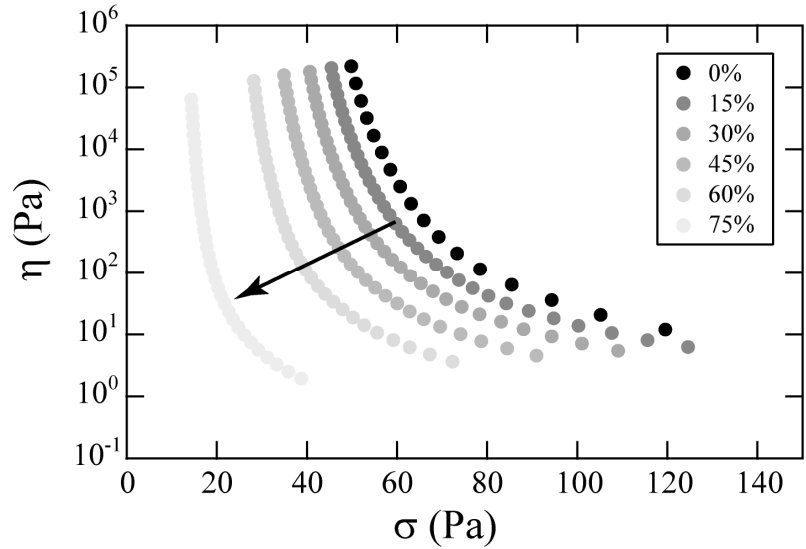

**Figure 3: Viscosity as a function of shear stress for different gel concentration (the percentage indicated here is the % of water in weight in the mixture). The black arrow indicates the evolution when diluting the gel.**

In Figure 3, we present the evolution of the fluid viscosity as a function of shear stress, for different gel-water mixtures. Note the continuous evolution of the rheological curves. In a previous study, we presented the evolution of such curves for the 10% gel, when the void fraction (number of small bubbles trapped in the fluid) increases (see Divoux et al. [2009], Fig.11). The evolution is similar, indicating that the presence of bubbles strongly influences the fluid rheology.

## References

----------------------------------------------------